


\def\drafttitle{global.tex}
\input lanlmac
\input amssym
\input epsf

\newcount\figno
\figno=0
\def\fig#1#2#3{
\par\begingroup\parindent=0pt\leftskip=1cm\rightskip=1cm\parindent=0pt
\baselineskip=13pt
\global\advance\figno by 1
\midinsert
\epsfxsize=#3
\centerline{\epsfbox{#2}}
\vskip 12pt
{\bf Fig. \the\figno:~~} #1 \par
\endinsert\endgroup\par
}
\def\figlabel#1{\xdef#1{\the\figno}}
\newdimen\tableauside\tableauside=1.0ex
\newdimen\tableaurule\tableaurule=0.4pt
\newdimen\tableaustep
\def\phantomhrule#1{\hbox{\vbox to0pt{\hrule height\tableaurule
width#1\vss}}}
\def\phantomvrule#1{\vbox{\hbox to0pt{\vrule width\tableaurule
height#1\hss}}}
\def\sqr{\vbox{%
  \phantomhrule\tableaustep

\hbox{\phantomvrule\tableaustep\kern\tableaustep\phantomvrule\tableaustep}%
  \hbox{\vbox{\phantomhrule\tableauside}\kern-\tableaurule}}}
\def\squares#1{\hbox{\count0=#1\noindent\loop\sqr
  \advance\count0 by-1 \ifnum\count0>0\repeat}}
\def\tableau#1{\vcenter{\offinterlineskip
  \tableaustep=\tableauside\advance\tableaustep by-\tableaurule
  \kern\normallineskip\hbox
    {\kern\normallineskip\vbox
      {\gettableau#1 0 }%
     \kern\normallineskip\kern\tableaurule}%
  \kern\normallineskip\kern\tableaurule}}
\def\gettableau#1 {\ifnum#1=0\let\next=\null\else
  \squares{#1}\let\next=\gettableau\fi\next}

\tableauside=1.0ex
\tableaurule=0.4pt


\def\th{\theta}

\def\Tr{{\rm Tr}}
\def\hf{{1\over 2}}
\def\qu{{1\over 4}}

\def\si{\sigma}
\def\Si{\Sigma}

\def\lap{\Delta}
\def\bra{\langle}
\def\ket{\rangle}
\def\lf{\left}
\def\ri{\right}
\def\riya{\rightarrow}

\def\la{\lambda}

\def\h#1{\widehat{#1}}

\def\Ga{\Gamma}
\def\al{\alpha}

\def\Om{\Omega}

\def\rt#1{\sqrt{#1}}

\def\sitarel#1#2{\mathrel{\mathop{\kern0pt #1}\limits_{#2}}}

\lref\SenVM{
  A.~Sen,
  ``Quantum Entropy Function from AdS$_2$/CFT$_1$ Correspondence,''
  Int.\ J.\ Mod.\ Phys.\  A {\bf 24}, 4225 (2009)
  [arXiv:0809.3304 [hep-th]].
}
\lref\DrukkerZQ{
  N.~Drukker, D.~J.~Gross and H.~Ooguri,
  ``Wilson loops and minimal surfaces,''
  Phys.\ Rev.\  D {\bf 60}, 125006 (1999)
  [arXiv:hep-th/9904191].
}
\lref\DrukkerGA{
  N.~Drukker,
  ``1/4 BPS circular loops, unstable world-sheet instantons and the matrix
  model,''
  JHEP {\bf 0609}, 004 (2006)
  [arXiv:hep-th/0605151].
}
\lref\MaldacenaIM{
  J.~M.~Maldacena,
  ``Wilson loops in large N field theories,''
  Phys.\ Rev.\ Lett.\  {\bf 80}, 4859 (1998)
  [arXiv:hep-th/9803002].
}
\lref\ReyIK{
  S.~J.~Rey and J.~T.~Yee,
  ``Macroscopic strings as heavy quarks in large N gauge theory and  anti-de
  Sitter supergravity,''
  Eur.\ Phys.\ J.\  C {\bf 22}, 379 (2001)
  [arXiv:hep-th/9803001].
}
\lref\PestunRZ{
  V.~Pestun,
  ``Localization of gauge theory on a four-sphere and supersymmetric Wilson
  loops,''
  arXiv:0712.2824 [hep-th].
}
\lref\GiombiEK{
  S.~Giombi and V.~Pestun,
  ``The 1/2 BPS 't Hooft loops in N=4 SYM as instantons in 2d Yang-Mills,''
  arXiv:0909.4272 [hep-th].
}
\lref\EricksonAF{
  J.~K.~Erickson, G.~W.~Semenoff and K.~Zarembo,
  ``Wilson loops in N = 4 supersymmetric Yang-Mills theory,''
  Nucl.\ Phys.\  B {\bf 582}, 155 (2000)
  [arXiv:hep-th/0003055].
}
\lref\OkuyamaZN{
  K.~Okuyama,
  ``N = 4 SYM on $R \times S^3$ and pp-wave,''
  JHEP {\bf 0211}, 043 (2002)
  [arXiv:hep-th/0207067].
}
\lref\NicolaiEK{
  H.~Nicolai, E.~Sezgin and Y.~Tanii,
  ``Conformally Invariant Supersymmetric Field Theories On 
  $S^p \times S^1$ And
  Super P-Branes,''
  Nucl.\ Phys.\  B {\bf 305}, 483 (1988).
}
\lref\DrukkerQR{
  N.~Drukker, S.~Giombi, R.~Ricci and D.~Trancanelli,
  ``Supersymmetric Wilson loops on $S^3$,''
  JHEP {\bf 0805}, 017 (2008)
  [arXiv:0711.3226 [hep-th]].
}
\lref\DymarskyVE{
  A.~Dymarsky, S.~S.~Gubser, Z.~Guralnik and J.~M.~Maldacena,
  ``Calibrated surfaces and supersymmetric Wilson loops,''
  JHEP {\bf 0609}, 057 (2006)
  [arXiv:hep-th/0604058].
}
\lref\tHooftHY{
  G.~'t Hooft,
  ``On The Phase Transition Towards Permanent Quark Confinement,''
  Nucl.\ Phys.\  B {\bf 138}, 1 (1978).
}
\lref\tHooftUJ{
  G.~'t Hooft,
  ``A Property Of Electric And Magnetic Flux In Nonabelian Gauge Theories,''
  Nucl.\ Phys.\  B {\bf 153}, 141 (1979).
}
\lref\GrossGK{
  D.~J.~Gross and H.~Ooguri,
  ``Aspects of large N gauge theory dynamics as seen by string theory,''
  Phys.\ Rev.\  D {\bf 58}, 106002 (1998)
  [arXiv:hep-th/9805129].
}
\lref\BerensteinIJ{
  D.~E.~Berenstein, R.~Corrado, W.~Fischler and J.~M.~Maldacena,
  ``The operator product expansion for Wilson loops and surfaces in the  large
  N limit,''
  Phys.\ Rev.\  D {\bf 59}, 105023 (1999)
  [arXiv:hep-th/9809188].
}
\lref\OlesenJI{
  P.~Olesen and K.~Zarembo,
  ``Phase transition in Wilson loop correlator from AdS/CFT correspondence,''
  arXiv:hep-th/0009210.
}
\lref\DrukkerRR{
  N.~Drukker and D.~J.~Gross,
  ``An exact prediction of N = 4 SUSYM theory for string theory,''
  J.\ Math.\ Phys.\  {\bf 42}, 2896 (2001)
  [arXiv:hep-th/0010274].
}
\lref\KimTD{
  H.~Kim, D.~K.~Park, S.~Tamarian and H.~J.~W.~Muller-Kirsten,
  ``Gross-Ooguri phase transition at zero and finite temperature: Two  circular Wilson loop case,''
  JHEP {\bf 0103}, 003 (2001)
  [arXiv:hep-th/0101235].
}
\lref\GorskyPC{
  A.~Gorsky, A.~Monin and A.~V.~Zayakin,
  ``Correlator of Wilson and 't Hooft Loops at Strong Coupling in
  N=4 SYM Theory,''
  Phys.\ Lett.\  B {\bf 679}, 529 (2009)
  [arXiv:0904.3665 [hep-th]].
}
\lref\ZaremboBU{
  K.~Zarembo,
  ``Wilson loop correlator in the AdS/CFT correspondence,''
  Phys.\ Lett.\  B {\bf 459}, 527 (1999)
  [arXiv:hep-th/9904149].
}
\lref\KimTD{
  H.~Kim, D.~K.~Park, S.~Tamarian and H.~J.~W.~Muller-Kirsten,
  ``Gross-Ooguri phase transition at zero and finite 
temperature: Two  circular
  Wilson loop case,''
  JHEP {\bf 0103}, 003 (2001)
  [arXiv:hep-th/0101235].
}

\Title{             
                                              }
{\vbox{
\centerline{Global AdS Picture of 1/2 BPS Wilson Loops}
}}

\vskip .2in

\centerline{Kazumi Okuyama}
\vskip5mm
\centerline{Department of Physics, Shinshu University}
\centerline{Matsumoto 390-8621, Japan}
\centerline{\tt kazumi@azusa.shinshu-u.ac.jp}
\vskip .2in

\vskip 3cm
\noindent

We study the holographic dual string configuration of
1/2 BPS circular Wilson loops in ${\cal N}=4$
super Yang-Mills theory by using
the global coordinate of AdS.
The dual string worldsheet is given by the
Poincar\'{e} disk $AdS_2$ sitting at a constant global time
slice of $AdS_5$. We also analyze the
correlator of
two concentric circular Wilson loops
from the global AdS perspective
and study the phase transition associated
with the instability of annulus worldsheet
connecting the two Wilson loops.

\Date{December 2009}

\vfill
\vfill
\newsec{Introduction}
In the AdS/CFT correspondence between
${\cal N}=4$ super Yang-Mills (SYM) and the type IIB string theory
on $AdS_5\times S^5$, supersymmetric Wilson loops
are interesting objects to study.
On the bulk gravity side, the expectation value of
such Wilson loops is obtained by computing
the minimal area of string worldsheet ending on the loop
at the boundary of AdS \refs{\MaldacenaIM,\ReyIK}.
The string worldsheet which is holographically
dual to 
a supersymmetric Wilson loop
is generally characterized as
a pseudo-holomorphic curve
on $AdS_5\times S^5$ \refs{\DrukkerQR,\DymarskyVE}.
However, it would be nice to have a more 
intuitive picture of dual worldsheet.

In the case of a 1/2 BPS circular Wilson loop,
the dual worldsheet was obtained in 
the Pincar\'{e} coordinate of $AdS_5$
\BerensteinIJ.
In this paper, we consider the worldsheet
dual of 1/2 BPS circular Wilson loops
using the global coordinate of $AdS_5$.
This global AdS description of
Wilson loop is closely related to
the radial quantization of ${\cal N}=4$ SYM,
since the ${\cal N}=4$ SYM defined on ${\Bbb R}\times S^3$
is dual to type IIB string theory
on global $AdS_5$. 
We find that the 
dual worldsheet of 1/2 BPS circular Wilson loop
is the Poincar\'{e} disk sitting at a fixed global time
of $AdS_5$.
Using this picture, we revisit the
holographic computation 
of the correlator of two concentric 1/2 BPS circular Wilson loops
studied in \OlesenJI.
As discussed in \OlesenJI,
when the ratio of radii of the two circles exceeds a certain
critical value, the worldsheet of annulus topology
in the bulk AdS ceases to exist,
and this leads to
an analogue of Gross-Ooguri phase transition \GrossGK.
We analyze the annulus worldsheet connecting
the two loops using the global coordinate of $AdS_5$.

This paper is organized as follows.
In section 2, we first consider the
radial quantization picture
of 1/2 BPS circular Wilson loops
in the SYM side.
Next we study the dual worldsheet
ending on the circular loop using the global coordinate
of $AdS_5$.
In section 3, we study the 
correlator of two concentric circular Wilson loops
from the global AdS perspective and 
consider the Gross-Ooguri type
transition
of this correlator.
In section 4, we discuss some interesting future directions.

\newsec{$1/2$ BPS Wilson Loops from Global AdS Perspective}
\subsec{Circular Wilson Loops in Radial Quantization}
In this paper we consider Wilson loops in
${\cal N}=4$ SYM on Euclidean signature space.
It is well-known that the 1/2 BPS Wilson loop in ${\cal N}=4$ SYM
is given by \MaldacenaIM
\eqn\maldacenaloop{
W(C)={1\over N}\Tr P\exp\left[\oint_C ds \Big(iA_\mu(x)\dot{x}^\mu(s)
+\theta^I\Phi_I(x)|\dot{x}(s)|\Big)\right]~,
}
where $C$ is a circle or a straight line on ${\Bbb R}^4$
in order to preserve 1/2 of supersymmetry.
$\th^I$ in \maldacenaloop\ is a constant unit 6-vector which specifies a
point on $S^5$.
Let us consider a circular Wilson loop $C$ with radius $a$
\eqn\circleC{
C: \quad x_1^2+x_2^2=a^2,\quad x_3=x_4=0~.
}
Here $x_\mu\,(\mu=1,..,4)$ denote the coordinate
of Euclidean ${\Bbb R}^4$.

In the radial quantization
with respect to the origin of ${\Bbb R}^4$,
we introduce the radial time $\tau$ as the log of
radial coordinate $r=\rt{x_\mu^2}$.
Then the coordinate $x_\mu$ of ${\Bbb R}^4$ is written as
\eqn\tauvsr{
x_\mu=e^\tau n_\mu~,
}
where $n_\mu$ is a unit 4-vector $n_\mu^2=1$ parametrizing $S^3$.
By the change of variable $r=e^\tau$,
the metric of
${\Bbb R}^4$ becomes conformally equivalent to the 
metric of ${\Bbb R}\times S^3$
\eqn\dsconf{
ds^2=dr^2+r^2d\Om_3^2=e^{2\tau}(d\tau^2+d\Om_3^2)~,
}
where $d\Om_3^2$ is the metric of unit 3-sphere.
In this radial quantization picture,
the circular loop $C$ in \circleC\ 
becomes a great circle of $S^3=\{n_\mu^2=1\}$,
and $C$ is sitting at the constant radial time $\tau$
\eqn\tauloga{
\tau=\log a,\quad n_1^2+n_2^2=1~.
}

In the radial quantization, a local operator ${\cal O}_{\Delta}$ inserted
at the origin of
${\Bbb R}^4$ corresponds to a state $|\lap\ket$
via the state-operator correspondence
\eqn\Ovslap{
|\lap\ket={\cal O}_{\lap}(0)|0\ket~.
} 
What does 
the circular Wilson loop centered at the origin
of ${\Bbb R}^4$ correspond to in the radial quantization? 
The above argument suggests that
the  circular Wilson loop $W(C)$ with radius $a$
corresponds to an operator $\h{W}_C(\tau)$
inserted on a fixed time slice $\tau=\log a$
of ${\Bbb R}\times S^3$,
and this operator $\h{W}_C(\tau)$
acts on the Hilbert space ${\cal H}=\{|\lap\ket\}$
of ${\cal N}=4$ SYM.
For instance, 
the OPE of a Wilson loop $W(C)$ and a local operator
${\cal O}_{\lap}$ is written as
\eqn\opeWO{
\bra W(C){\cal O}_{\lap}(0)\ket=\bra 0|\h{W}_C(\tau)|\lap\ket~.
}
Similarly, the correlator of $n$ 
concentric 1/2 BPS circular Wilson loops
is written as
\eqn\radialpicture{
\bra W(C_1)\cdots W(C_n)\ket=\bra 0|{\bf T}\big[\h{W}_{C_1}(\tau_1)\cdots
\h{W}_{C_n}(\tau_n)\big]|0\ket~,
}
where ${\bf T}$ denotes the time ordering
with respect to the radial time $\tau$ 
(or the radial ordering in the original ${\Bbb R}^4$ picture)
and $\tau_i\,(i=1,..,n)$ is related to the radius $a_i$ of loop $C_i$ by
$\tau_i=\log a_i$.

\subsec{Holographic Dual of Circular Wilson Loops}
The gravity dual of Wilson loop in ${\cal N}=4$
SYM is given by the string worldsheet in $AdS_5\times S^5$
bounded by the loop at the boundary of $AdS_5$
\refs{\MaldacenaIM,\ReyIK}.
In the case of 1/2 BPS circular Wilson loop,
the dual string worldsheet
was obtained as a minimal surface
in $AdS_5$ which minimizes the Nambu-Goto action.
In the Poincar\'{e} coordinate of (Euclidean) $AdS_5$
\eqn\AdSPoincare{
ds^2_{AdS_5}={dz^2+dx_\mu dx^\mu \over z^2}~,
}
the dual string worldsheet of circular Wilson loop $C$
\circleC\ is given by \BerensteinIJ
\eqn\WSinPoincare{
\Si: \quad z^2+x_1^2+x_2^2=a^2,\quad x_3=x_4=0~.
}
One can easily see that this surface
$\Si$ ends on the 
loop $C$ \circleC\ at the boundary $z=0$ of $AdS_5$.

What does this surface $\Si$ look like
in the global coordinate of
$AdS_5$? The metric of $AdS_5$ in the global coordinate
is given by
\eqn\globalmetric{
ds^2_{AdS_5}=\cosh^2\!\rho\, d\tau^2+d\rho^2+
\sinh^2\!\rho\, d\Om^2_3~.
}
To see the relation between the Poincar\'{e}
coordinate and the global coordinate,
it is convenient to view $AdS_5$
as a hypersurface in ${\Bbb R}^{1,5}$
\eqn\AdSinY{
\eta_{ab}Y^aY^b=-1~,
}
where 
$Y^a\,(a=0,...,5)$ denotes the coordinate
of ${\Bbb R}^{1,5}$ with metric $\eta_{ab}=(-+++++)$.
Above, we have set the AdS radius $R_{AdS}=1$ for simplicity.
In terms of the Poincar\'{e} coordinate $(z,x^\mu)$,
$Y^a$ is given by
\eqn\YainPoincare{
Y^0+Y^5=z+{x_\mu^2\over z},\quad
Y^0-Y^5={1\over z},\quad
Y^\mu={x^\mu\over z}~~(\mu=1,...,4)~.
}
On the other hand, 
$Y^a$ is written in the global coordinate as
\eqn\Yainglobal{
Y^0\pm Y^5=e^{\pm\tau}\cosh\rho,\quad
Y^\mu=n^\mu\sinh\rho,
}
where $n^\mu$ is a unit 4-vector $n_\mu^2=1$ parametrizing $S^3$.
From \YainPoincare\ and \Yainglobal, the Poincar\'{e} coordinate $(x,x^\mu)$
and the global coordinate $(\tau,\rho,n^\mu)$ are related as
\eqn\PvsGtrf{
z={e^\tau\over\cosh\rho}~,\quad
x^\mu=e^\tau n^\mu\tanh\rho~.
}
One can check that the metric of $AdS_5$ in Poincar\'{e} coordinate
\AdSPoincare\ becomes the metric in the global coordinate
\globalmetric\ under
this change of variables \PvsGtrf.

In the global coordinate, 
the boundary of $AdS_5$ is located at $\rho=\infty$.
At the boundary of $AdS_5$, $(z,x^\mu)$ in \PvsGtrf\
becomes
\eqn\Pzxbdry{
z\riya 0,\quad
x^\mu\riya e^\tau n^\mu\quad(\rho\riya\infty)~.
}
By comparing \tauvsr\ and \Pzxbdry,
we find that the radial time $\tau$
of ${\Bbb R}\times S^3$
is identified with the global time $\tau$
of $AdS_5$ at the boundary $\rho=\infty$.
This is consistent with the fact
that the ${\cal N}=4$ SYM defined on ${\Bbb R}\times S^3$
is dual to the type IIB string theory on global $AdS_5$.

Now let us rewrite the minimal surface $\Si$ \WSinPoincare\
using the global coordinate.
It turns out that the equation of $\Si$
in the global coordinate is simply given by the same
equation as the circular loop $C$
in the radial quantization
picture \tauloga
\eqn\Siintaun{
\tau=\log a,\quad n_1^2+n_2^2=1~.
}
For this configuration of worldsheet $\Si$,
the unit 4-vector $n^\mu$ in \Yainglobal\ can be taken as
\eqn\ninphi{
(n_1,n_2,n_3,n_4)=(\cos\phi,\sin\phi,0,0)~.
}
Plugging \Siintaun\ and \ninphi\ into the relation
between Poincar\'{e} and global coordinates \PvsGtrf,
we get
\eqn\Sigmainglobal{
z={a\over\cosh\rho},\quad
x_1+ix_2=ae^{i\phi}\tanh\rho~.
}
We can easily see that $(z,x_1,x_2)$ in \Sigmainglobal\ satisfy
the equation  of $\Si$ in the Poinca\'{e}
coordinate \WSinPoincare. Therefore, we
conclude that \Siintaun\ is the equation of minimal surface $\Si$ 
written in the global coordinate.
We can also show
that the metric on $\Si$ induced from the global $AdS_5$ metric
\globalmetric\
is the metric of $AdS_2$
\eqn\dsonSigma{
ds^2_{\Si}
=d\rho^2+\sinh^2\!\rho\,d\phi^2~.
}
To summarize, the minimal surface $\Si$ in the global coordinate
is a
Poincar\'{e} disk $AdS_2$
parametrized by $\rho\in[0,\infty]$
and $\phi\in[0,2\pi]$ and $\Si$ is 
sitting at a constant time slice $\tau=\log a$
of global $AdS_5$
(see Fig. 1). Note that $\phi$ in \ninphi\
is the angle coordinate of the great circle $S^1=\{n_1^2+n_2^2=1\}$
on the 3-sphere $S^3=\{n_\mu^2=1\}$.

\fig{The string worldsheet $\Si$ dual to a 1/2 BPS circular Wilson loop 
$W(C)$ is the Poincar\'{e} disk $AdS_2$ bounded by the loop $C$.
$\Si$ is sitting at a constant time slice of global $AdS_5$.
}{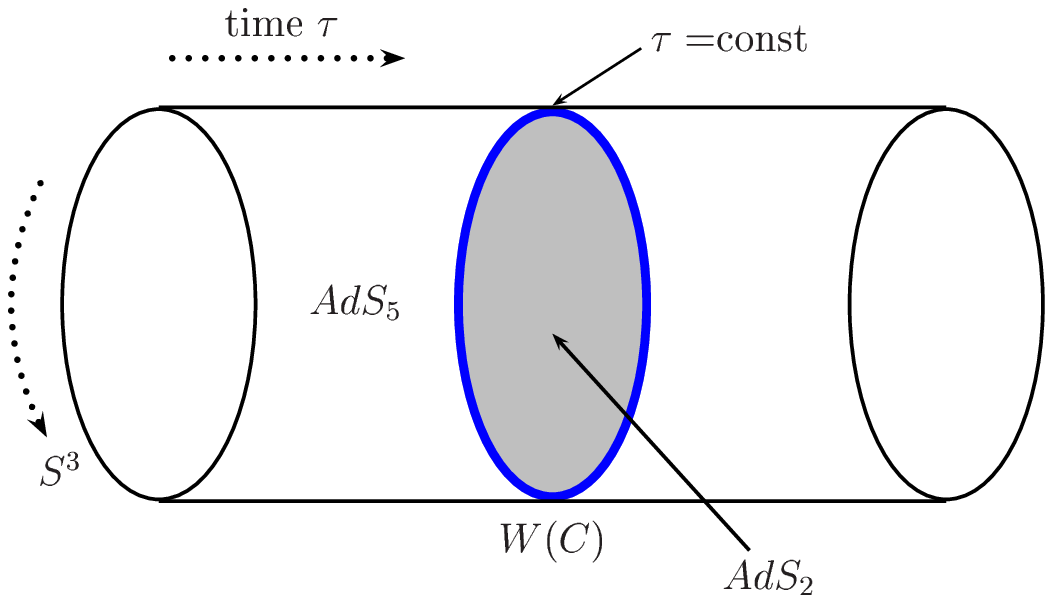}{6.5cm}
\subsec{Nambu-Goto Action in Global Coordinate}
In this subsection we consider the
equation of motion
of Nambu-Goto string
using the global coordinate.
The Nambu-Goto action in $AdS_5$ is given by
\eqn\NGaction{
S={\sqrt{\la}\over2\pi}\int d^2\si\rt{\det G}~,
}
where $G$ is the worldsheet metric induced from
the global $AdS_5$ \globalmetric.
Here we used the fact that the string tension in the unit of AdS
radius is related to the 't Hooft coupling $\la=g_{\rm YM}^2N$
of ${\cal N}=4$ SYM as
\eqn\TFvsla{
T_{F1}R_{AdS}^2={R_{AdS}^2\over2\pi\al'}={\sqrt{\la}\over2\pi}~.
}
In order to analyze the configuration of string worldsheet
in the global coordinate of $AdS_5$ \globalmetric,
we make an ansatz
\eqn\ansatzsheet{
\rho=\si_0,\quad \phi=\si_1,\quad
\tau=\tau(\si_0)=\tau(\rho)~,
}
where $(\si_0,\si_1)$ is the two-dimensional worldsheet coordinate.
As in the previous subsection, $\phi$ denotes the angular coordinate
of a great circle $S^1\subset S^3$ inside $AdS_5$. 
For this configuration \ansatzsheet,
the worldsheet metric induced from  $AdS_5$ \globalmetric\  
is given by
\eqn\indmetFone{
ds^2_{F1}=\left[\cosh^2\!\rho\left({d\tau\over d\rho}\right)^2+1\right]
d\rho^2+\sinh^2\!\rho \,d\phi^2~,
}
and the Nambu-Goto action \NGaction\ becomes
\eqn\NGintau{
S=
{\sqrt{\la}\over2\pi}\int d\phi d\rho
\,\sinh\!\rho\rt{\cosh^2\!\rho\left({d\tau\over d\rho}\right)^2+1}~.
}
The equation of motion following
from this action is
\eqn\eomtau{
{d\over d\rho}\left({\sinh\rho\cosh^2\rho{d\tau\over d\rho}\over
\rt{\cosh^2\!\rho\left({d\tau\over d\rho}\right)^2+1}}\right)
=0~.
}
Clearly, ${d\tau\over d\rho}=0$ is a solution of this equation
and the induced metric \indmetFone\ for the ${d\tau\over d\rho}=0$ case is 
\eqn\AdstwoWS{
ds^2_{F1}=d\rho^2+\sinh^2\!\rho\,d\phi^2.
}
This is nothing but the metric of $AdS_2$.
This analysis gives a direct check that the Poincar\'{e} disk sitting at
$\tau=\log a$ is a minimal surface.

The on-shell action
of the minimal surface has a divergence
coming from the large $\rho$ region, and
we need to regularize it by introducing the IR cut-off $\rho_0$
in the $\rho$-integral \NGintau
\eqn\loopaction{
S
={\sqrt{\la}\over2\pi}\int_0^{2\pi}d\phi\int_0^{\rho_0}d\rho 
\sinh\rho 
=\rt{\la}(\cosh\rho_0-1)~.
}
The first term $\rt{\la}\cosh\rho_0$ in \loopaction\
diverges in the limit $\rho_0\riya\infty$, and this can be
removed by adding a boundary term to the action\foot{This procedure 
is reminiscent of the quantum entropy function formalism of Sen \SenVM.} 
\refs{\MaldacenaIM,\DrukkerZQ}.
After removing this term, we get the regularized action
\eqn\loopregS{
S_{\rm reg}=-\rt{\la}~,
}
which gives the leading large $\lambda$ behavior of 
Wilson loop expectation value
\eqn\loopexp{
\bra W(C)\ket \approx e^{-S_{\rm reg}}=e^{\rt{\la}}~.
}
This large $\la$ behavior of $\bra W(C)\ket$
is reproduced by the exact computation in the ${\cal N}=4$
SYM side using the localization of path integral to a Gaussian matrix model
\refs{\EricksonAF,\DrukkerRR,\PestunRZ}.

\newsec{Two Concentric Circular Wilson Loops}
In this section, we study the correlator of
two concentric circular Wilson loops
from the holographic dual viewpoint.
This problem is studied in \OlesenJI\foot{The 
holographic dual of two Wilson loops 
with equal radii separated in the $x_3$-direction
was studied in \refs{\ZaremboBU,\KimTD}.}
and it is found that as we vary the ratio ${a_2\over a_1}$
of the radii $a_1,a_2$ of two circles there is a
phase transition similar to that found
by Gross and Ooguri \GrossGK.
We revisit this problem using the global coordinate of
$AdS_5$. As we will see below, the use of global coordinate enables
us to visualize the dual worldsheet very clearly.

In the radial quantization picture,
the correlator $\bra W(C_1)W(C_2)\ket$
of two circular Wilson loops is written as the two operator insertions 
at $\tau_i=\log a_i\,(i=1,2)$
\eqn\twoloopWhat{
\bra W(C_1)W(C_2)\ket=
\bra 0|\h{W}_{C_2}(\tau_2)\h{W}_{C_1}(\tau_1)|0\ket~.
}
Here we have assumed $\tau_2>\tau_1$ without loss of generality.
The two Wilson loops are separated in the $\tau$-direction
by the amount $\tau_0$
\eqn\tauzerodef{
\tau_0=\tau_2-\tau_1=\log{a_2\over a_1}~.
}
This correlator
has a connected part and a disconnected part
\eqn\WWcorr{
\bra W(C_1)W(C_2)\ket=
\bra W(C_1)W(C_2)\ket_{\rm conn}
+\bra W(C_1)\ket\bra W(C_2)\ket~.
}

\fig{(A) The worldsheet of annulus topology connecting
the two loops $C_1,C_2$ contributes to the connected
part of the correlator of two Wilson loops 
$\bra W(C_1)W(C_2)\ket_{\rm conn}$.
The minimal surface of annulus topology does not exist
if $\tau_0$ is larger than some critical value $\tau_c$.
(B) The two disconnected
disk worldsheets correspond to the disconnected
part of Wilson loop correlator $\bra W(C_1)\ket\bra W(C_2)\ket$. 
}{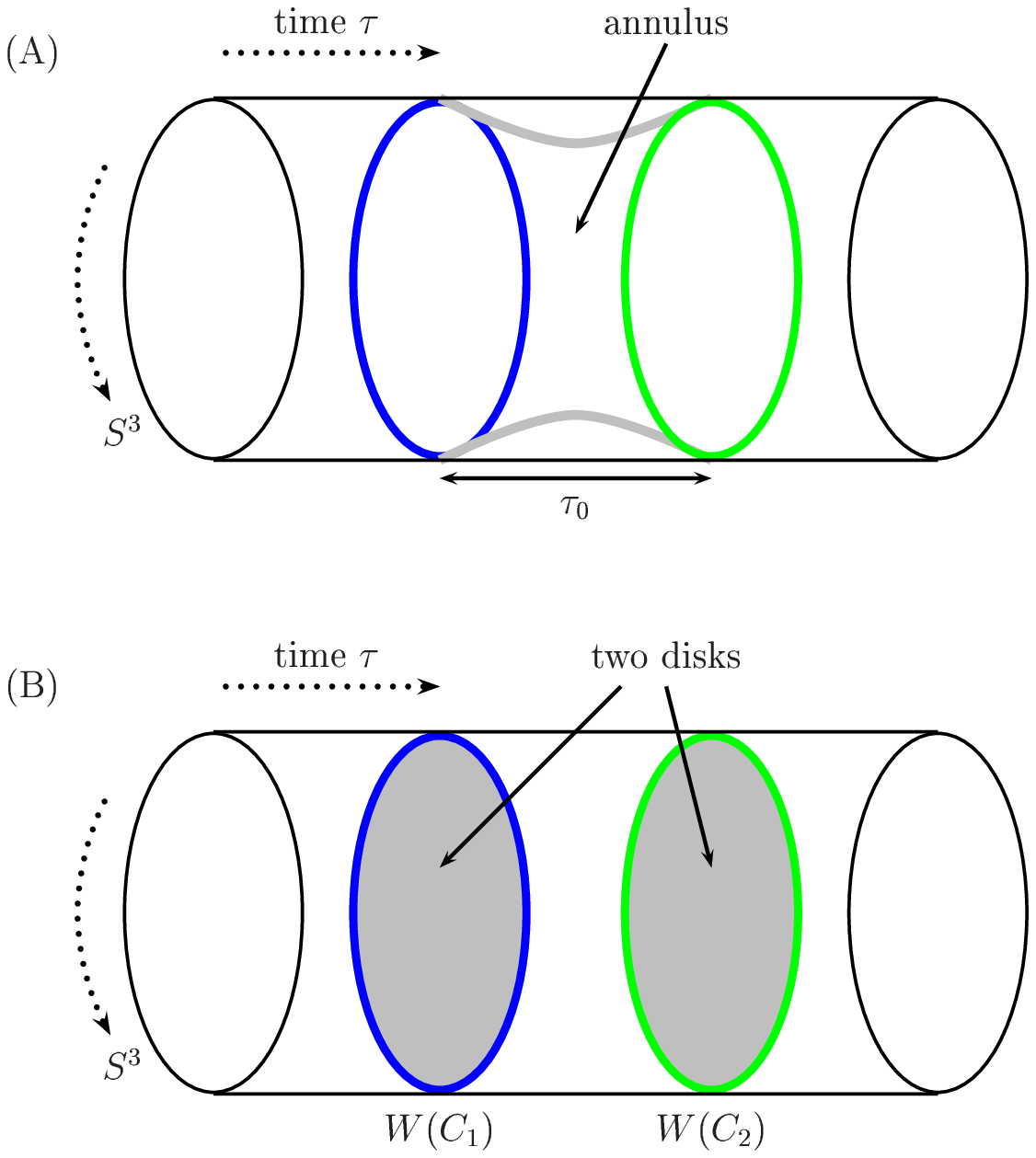}{7cm}

In the large 't Hooft coupling regime, the
correlator $\bra W(C_1)W(C_2)\ket$ has a holographic dual description
as a string worldsheet having the two circles
$C_1$ and $C_2$ as boundaries. In the global $AdS_5$ picture,
the disconnected part corresponds to the two
Poincar\'{e} disks sitting at $\tau=\tau_1$ and $\tau=\tau_2$,
and the connected part corresponds to the annulus worldsheet
connecting $C_1$ and $C_2$ (see Fig. 2).
Then the two-loop correlator \WWcorr\ in the large $\la$
regime is approximately given by
the regularized Nambu-Goto action
\eqn\WinSregann{\eqalign{
&\bra W(C_1)W(C_2)\ket_{\rm conn}\approx
e^{-S_{\rm reg}({\rm annulus})}~,\cr
&\bra W(C_1)\ket\bra W(C_2)\ket\approx
e^{-2S_{\rm reg}({\rm disk})}=e^{2\rt{\la}}~.
}}

In the following we study the annulus worldsheet connecting the two circles
using the global $AdS_5$ coordinate.
We make the same ansatz \ansatzsheet\ for the string configuration
as in the disk case,
hence the equation of motion
we should solve is \eomtau.
This equation \eomtau\ can be integrated once
\eqn\Ltauconst{
{\sinh\rho\cosh^2\rho {d\tau\over d\rho}\over\rt{\cosh^2\rho
({d\tau\over d\rho})^2+1}}={\rm const}.
}
Using the translation invariance in the $\tau$-direction,
we can set the locations of two loops to be
at $\tau_1=-\hf \tau_0$ and $\tau_2=\hf\tau_0$.
In order for the annulus worldsheet to
end on the two circles at the boundary of $AdS_5$,
$\tau(\rho)$ should satisfy the following boundary condition
\eqn\bdrytau{
\lim_{\rho\riya\infty}\tau(\rho)=\pm{\tau_0\over2}~.
}
Also, we require that $\tau(\rho)$ is vertical at the turning point
$\rho=\rho_{\rm min}$ (see Fig. 2(A))
\eqn\turningtau{
\lim_{\rho\riya\rho_{\rm min}}{d\tau\over d\rho}=\infty~.
}
From this condition the constant on the right hand side of 
\Ltauconst\ is fixed as
\eqn\constLtau{
{\sinh\rho\cosh^2\rho  {d\tau\over d\rho}\over\rt{\cosh^2\rho
( {d\tau\over d\rho})^2+1}}=\cosh\rho_{\rm min}\sinh\rho_{\rm min}~.
} 
From this equation we get
\eqn\delrhotau{
{d\tau\over d\rho}=\pm 
{\sinh^22\rho_{\rm min}\over\cosh\rho\rt{\sinh^22\rho
-\sinh^22\rho_{\rm min}}}~.
}
The plus sign of \delrhotau\ represents the right half of annulus
$0\leq\tau\leq\hf \tau_0$, 
and the minus sign corresponds to the left half
$-\hf \tau_0\leq\tau\leq0$ (see Fig. 2(A)). 
On the right half of annulus,
the equation \delrhotau\ is solved as
\eqn\tauvsrho{
\tau(\rho)=\int_{\rho_{\rm min}}^\rho
{d\rho\over\cosh\rho}{\sinh2\rho_{\rm min}\over\rt{\sinh^22\rho
-\sinh^22\rho_{\rm min}}}~.
}
The boundary condition \bdrytau\ leads to the relation
between $\tau_0$ and $\rho_{\rm min}$
\eqn\tauzeroint{
\tau_0=2\int_{\rho_{\rm min}}^\infty
{d\rho\over\cosh\rho}{\sinh2\rho_{\rm min}\over\rt{\sinh^22\rho
-\sinh^22\rho_{\rm min}}}\equiv f(\rho_{\rm min})~.
}
By performing the change of variable
$t={\sinh\rho_{\rm min}\over \sinh\rho}$,
the function $f(\rho_{\rm min})$
in \tauzeroint\ is written as\foot{Our function
$f(\rho_{\rm min})$ corresponds to the function $2F(ka)$ in
\refs{\ZaremboBU,\OlesenJI}.} 
\eqn\tauasKPI{
f(\rho_{\rm min})=2\cosh\rho_{\rm min}\lf[K(i\coth\rho_{\rm min})
-\Pi\lf(-{1\over\sinh^{2}\rho_{\rm min}},i\coth\rho_{\rm min}\ri)\ri]~,
}
where $K(k)$ and $\Pi(n,k)$ denote the elliptic integrals of
the first and third kind, respectively
\eqn\KPidef{
K(k)=\int_0^1{dt\over\rt{(1-t^2)(1-k^2t^2)}}~,~~
\Pi(n,k)=\int_0^1{dt\over(1-nt^2)\rt{(1-t^2)(1-k^2t^2)}}~.
}

The function $f(\rho_{\rm min})$
vanishes in the limit $\rho_{\rm min}\riya 0,\infty$ as
\eqn\frholim{\eqalign{
f(\rho_{\rm min})&
\sim -2\rho_{\rm min}\log \rho_{\rm min}\quad(\rho_{\rm min}\riya0),\cr
f(\rho_{\rm min})&\sim {4\rt{2\pi^3}\over\Ga\left(\qu\right)^2}
e^{-\rho_{\rm min}}\quad(\rho_{\rm min}\riya\infty),
}}
and $f(\rho_{\rm min})$
has a single maximum at some finite $\rho_{\rm min}=\rho_c$
(see Fig. 3). As discussed in \OlesenJI,
the existence of a maximal value $\tau_c=f(\rho_c)$
of the function $f(\rho_{\rm min})$
means that the annulus worldsheet 
ceases to exist when
the separation $\tau_0$ between the two loops is
larger than $\tau_c$, and the annulus configuration
exists only when $0\leq\tau_0\leq\tau_c$.
This is analogous to the phase transition discussed by Gross and Ooguri
\GrossGK. 
\fig{This is a plot 
of $f(\rho_{\rm min})$ 
as a function of $b={1\over\sinh\rho_{\rm min}}$.
The function $f(\rho_{\rm min})$ has a single maximum with maximal value
$\tau_c(\sim0.68)$.
}{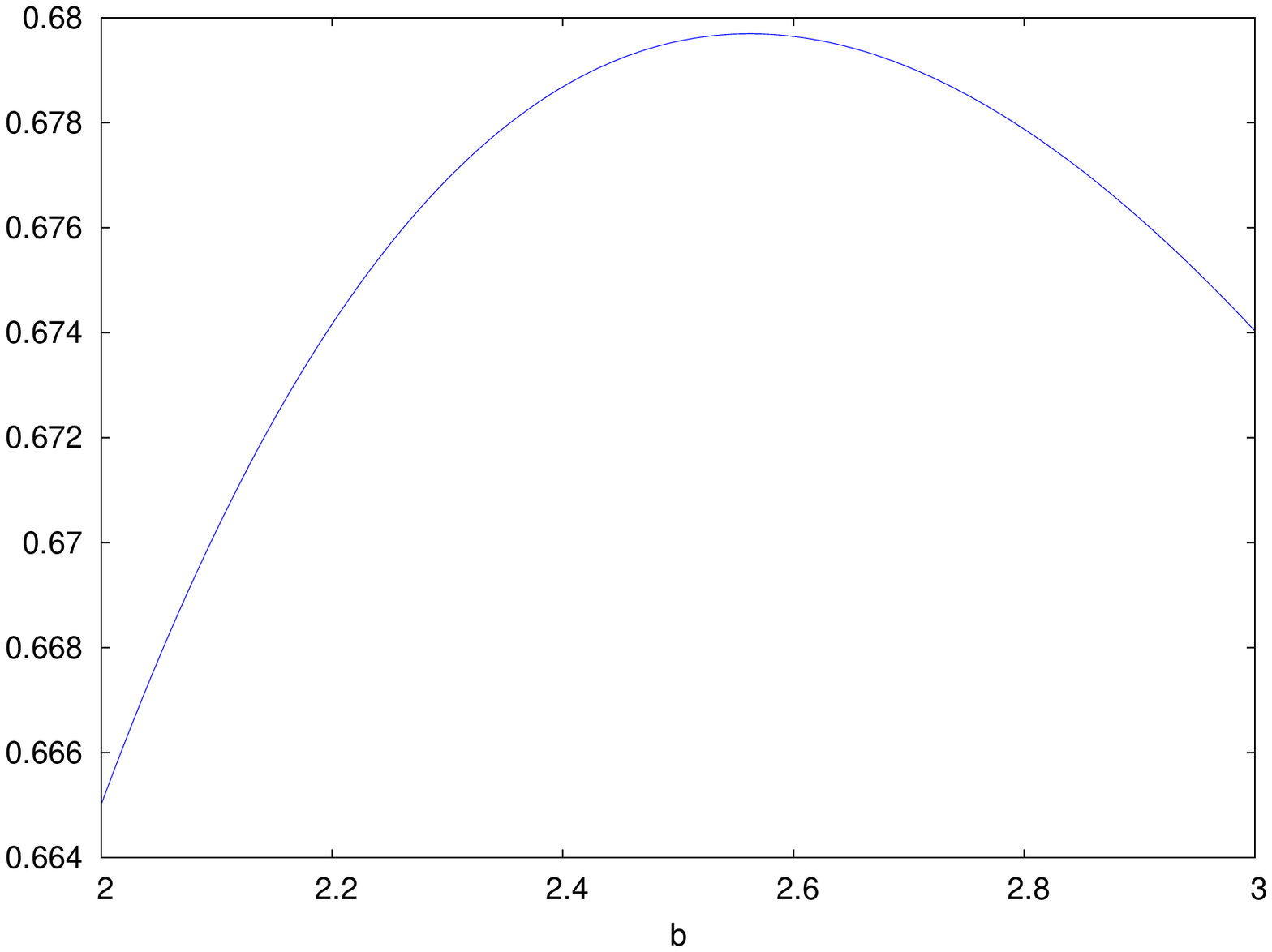}{5cm}

Next we consider the on-shell action 
for this annulus configuration.
Plugging ${d\tau\over d\rho}$ in \delrhotau\ into 
the action \NGintau, we find
\eqn\Swsfortwo{\eqalign{
S&=2\rt{\la}\int_{\rho_{\rm min}}^\infty d\rho\sinh\rho\rt{
\cosh^2\rho\left({d\tau\over d\rho}\right)^2+1}\cr
&=\rt{\la}\int_{\rho_{\rm min}}^\infty {d\rho\over\cosh\rho}
{\sinh^22\rho\over\rt{\sinh^22\rho-\sinh^22\rho_{\rm min}}}~.
}}
The factor of $2$ in the first line
of \Swsfortwo\ comes from the fact that
there are two branches ($\pm$ in \delrhotau)
for $\tau(\rho)$. 
To evaluate this integral, we separate it into two parts
\eqn\Sonwtwo{\eqalign{
S&=S^{(1)}+S^{(2)}~,\cr
S^{(1)}&=\rt{\la}\int_{\rho_{\rm min}}^\infty {d\rho\over\cosh\rho}
\rt{\sinh^22\rho-\sinh^22\rho_{\rm min}}~,\cr
S^{(2)}&=\rt{\la}\int_{\rho_{\rm min}}^\infty {d\rho\over\cosh\rho}
{\sinh^22\rho_{\rm min}\over \rt{\sinh^22\rho-\sinh^22\rho_{\rm min}}}~.
}}
For $S^{(1)}$, we 
further separate it into a divergent part $S^{(1)}_{\rm div}$
and a finite part $S^{(1)}_{\rm fin}$
\eqn\Soneexpand{\eqalign{
S^{(1)}_{\rm div}&=\rt{\la}
\int_{\rho_{\rm min}}^{\rho_0} {d\rho\over\cosh\rho}\sinh2\rho
=2\rt{\la}(\cosh\rho_0-\cosh\rho_{\rm min}),\cr
S^{(1)}_{\rm fin}&=
\rt{\la}\int_{\rho_{\rm min}}^\infty 
{d\rho\over\cosh\rho}\lf(\rt{\sinh^22\rho-\sinh^22\rho_{\rm min}}
-\sinh2\rho\ri).
}}
Here we have introduced the IR cut-off $\rho_0$ in $S^{(1)}_{\rm div}$
as before. Note that $S^{(1)}_{\rm fin}$ is negative.
$S^{(2)}$ in \Sonwtwo\ is proportional to the function
$f(\rho_{\rm min})$ in \tauzeroint
\eqn\Stwoinf{
S^{(2)}=\hf \rt{\la}f(\rho_{\rm min})\sinh2\rho_{\rm min}~.
}

After removing 
the divergent piece $2\rt{\la}\cosh\rho_0$ in $S^{(1)}_{\rm div}$,
the regularized action for the annulus worldsheet is found to be 
\foot{The first and the second terms of $S_{\rm reg}({\rm annulus})$
in (3.19) have a simple interpretation: The first term is the regularized area
of two disks at $\tau=\pm\hf\tau_0$,
where each disk has a hole of size $\rho_{\rm min}$
at its center. The second term represents the area of
a cylinder $I\times S^1_{\phi}$ sitting at $\rho=\rho_{\rm min}$.
Here $I$ denotes the interval of
$\tau$ with length $\tau_0=f(\rho_{\rm min})$.
$S^{(1)}_{\rm fin}$ represents the
correction
to this crude approximation
that the annulus worldsheet is made of two disks with holes
connected by a cylinder of radius $\rho_{\rm min}$ and
length $\tau_0$.
}
\eqn\Sregann{
S_{\rm reg}({\rm annulus})=-2\rt{\la}\cosh\rho_{\rm min}+
\hf \rt{\la}f(\rho_{\rm min})\sinh2\rho_{\rm min}+S^{(1)}_{\rm fin}
~.
}
This is written as a function in $\rho_{\rm min}$,
but we would like to study the behavior of this
action as a function of $\tau_0$ in the region $0<\tau_0<\tau_c$ where
the annulus worldsheet exists.
When rewriting this action as a function of $\tau_0$, 
we should be careful about the
fact that there are 
two solutions $\rho_{\rm min}^{(1)},\rho_{\rm min}^{(2)}$
for the equation
$\tau_0=f(\rho_{\rm min})$ since the function $f(\rho_{\rm min})$
takes a maximum value $\tau_c$
at $\rho_{\rm min}=\rho_c$.
We assume that $\rho_{\rm min}^{(1)}<\rho_c<\rho_{\rm min}^{(2)}$.
The existence of two solutions for $\tau_0=f(\rho_{\rm min})$
implies that
given a separation of two loops $\tau_0$
there are two annulus configurations corresponding
to $\rho_{\rm min}^{(1,2)}$.

We are interested in the difference
$\lap S$ between the annulus action \Sregann\
and the action of
two disks
$2S_{\rm reg}({\rm disk})=-2\rt{\la}$
\eqn\Slapad{
\lap S=2\rt{\la}(1-\cosh\rho_{\rm min})+
\hf \rt{\la}f(\rho_{\rm min})\sinh2\rho_{\rm min}+S^{(1)}_{\rm fin}~.
}
As analyzed numerically in \OlesenJI, 
the annulus action for the second branch 
$\rho_{\rm min}^{(2)}>\rho_c$ is smaller than
that of the first branch $\rho_{\rm min}^{(1)}<\rho_c$.
Therefore,
the second branch $\rho_{\rm min}>\rho_c$ gives the dominant 
contribution
to the connected part of correlator $\bra W(C_1)W(C_2)\ket_{\rm conn}$.
In particular, $\lap S$ in \Slapad\ becomes
arbitrarily negative as $\rho_{\rm min}\riya\infty$,
which corresponds
to the coincident limit of two loops $\tau_0\riya0$.
This can be seen as follows.
From \frholim, the large $\rho_{\rm min}$ limit of
$\lap S$ in \Slapad\ is written as
\eqn\largerhoS{
\lap S\sim \rt{\la}\left[-1+{\rt{2\pi^3}\over\Ga\left(\qu\right)^2}\right]
e^{\rho_{\rm min}}+S^{(1)}_{\rm fin}~.
}
Since the coefficient of $e^{\rho_{\rm min}}$ is negative
and also $S^{(1)}_{\rm fin}<0$, $\lap S\riya -\infty$ as
$\rho_{\rm min}\riya\infty$. As we increase $\tau_0$ from 0,
$\lap S$ becomes positive at certain value $\tau_c'$,
which occurs before the disappearance of annulus
solution, {\it i.e.}
$\tau_c'<\tau_c$. At $\tau_0=\tau_c'$ the annulus
configuration becomes unstable and 
collapses to two disks \GrossGK. 

Next we consider the $\rho_{\rm min}<\rho_c$ region.
The annulus action for this branch is always
larger than that of two disks \OlesenJI.
We can easily see that the annulus action 
\Sregann\ 
reduces to the action of two disks in the limit
$\rho_{\rm min}\riya0$
\eqn\Sannlim{
\lim_{\rho_{\rm min}\riya0}\lap S=0~.
}
Let us see that $\lap S>0$
for small $\rho_{\rm min}\ll1$.
Using \frholim\ and $S^{(1)}_{\rm fin}\sim\rt{\la}
\rho_{\rm min}^2\log\rho_{\rm min}$
we find
\eqn\diffSandisk{
\lap S\sim -\rt{\la}\rho_{\rm min}^2\log\rho_{\rm min}~,
}
which is indeed positive.

\newsec{Discussion}
In this paper we have considered the 
minimal surface in the global $AdS_5$ which is 
dual to 1/2 BPS circular Wilson loops.
The dual worldsheet is given by 
the Pincar\'{e} disk at a fixed global time.
We also revisited the computation of
the correlator of two concentric 
Wilson loops using the global coordinate of $AdS_5$.
The annulus worldsheet connecting the two loops 
exists only when the separation $\tau_0$
between the two loops is less than some critical value $\tau_c$,
and it ceases to exist
when $\tau_0>\tau_c$.
As argued in \OlesenJI, 
this phase transition is different from
the point where
the annulus and two disks change dominance.
It would be nice to understand the
physical picture of this transition more clearly.

It would be interesting to study the
radial quantization picture of
less supersymmetric Wilson loops or other
operators such as surface operators.
Another interesting direction to study is the commutation
relation between a 't Hooft loop and a Wilson loop 
from the bulk AdS viewpoint.
For a circular 't Hooft loop $T(C)$, the dual object is
a D1-brane with $AdS_2$ worldvolume
sitting at a constant global time.
In the radial quantization picture, the 't Hooft loop
$T(C)$ is represented by an operator $\h{T}_C(\tau)$
and this operator does not commute with
the Wilson loop operator $\h{W}_C(\tau)$ \refs{\tHooftHY,\tHooftUJ}
\eqn\thooftalg{
\h{T}_{C_1}(\tau_1)\h{W}_{C_2}(\tau_2)=e^{{2\pi i\over N}\ell(C_1,C_2)}
\h{W}_{C_2}(\tau_2)\h{T}_{C_1}(\tau_1)~,
}
where $\ell(C_1,C_2)$ is the linking number of two loops inside $S^3$.
Recently, the correlator of 
a 't Hooft loop $T(C_1)$ and a Wilson loop $W(C_2)$
in ${\cal N}=4$ SYM
is computed
by using the localization of path integral to
the two-dimensional Yang-Mills theory on $S^2$ \GiombiEK.
This localization technique works
when $C_1$ and $C_2$ are linked on $S^3$.
It would be interesting to understand the algebra
\thooftalg\ both from the gauge theory side and from
the gravity side \foot{See \GorskyPC\ for some
computation of the correlator of
concentric circular 't Hooft and Wilson loops 
on the gravity side.}.
On the gravity side, the algebra
\thooftalg\ corresponds to exchanging the D1-brane 
at $\tau=\tau_1$ and the fundamental string 
at $\tau=\tau_2$.
It would be nice to see how the phase
in \thooftalg\ appears in this change of ordering of
D1-brane and F1-brane along the $\tau$-direction in global $AdS_5$.
\vskip7mm
\noindent
\centerline{\bf Acknowledgment}
This work is supported in part by
MEXT Grant-in-Aid for Scientific Research \#19740135.

\listrefs
\bye